\documentclass{article}

\usepackage{arxiv}

\usepackage[utf8]{inputenc} 
\usepackage[T1]{fontenc}    
\usepackage{hyperref}       
\usepackage{url}            
\usepackage{booktabs}       
\usepackage{amsfonts}       
\usepackage{nicefrac}       
\usepackage{microtype}      
\usepackage{lipsum}
\usepackage{graphicx}
\usepackage{natbib}
\graphicspath{ {./figs/} }
\usepackage{array}
\newcolumntype{P}[1]{>{\raggedright\arraybackslash}p{#1}}
\usepackage{algorithm}
\usepackage{algorithmic}
\usepackage[capitalise]{cleveref}
\usepackage{todonotes}
\usepackage{pifont} 
\usepackage[nohyperlinks, nolist]{acronym} 
\usepackage{subcaption}
\usepackage{tcolorbox} 
\usepackage{makecell} 

\usepackage{ifthen}
\newboolean{singlecolumn}
\setboolean{singlecolumn}{true} 

\newcommand{\art}[1]{Art.~#1}

\usepackage{newfloat}
\usepackage{listings}
\usepackage{booktabs}
\DeclareCaptionStyle{ruled}{labelfont=normalfont,labelsep=colon,strut=off} 
\floatstyle{ruled}
\newfloat{listing}{tb}{lst}{}
\floatname{listing}{Listing}
\author{
  Anton Hummel \\
  XITASO GmbH \& University of Bayreuth\\
  Austraße 35, 86153 Augsburg, Germany\\
  \texttt{anton.hummel@xitaso.com} \\
   \And
  Håkan Burden \\
  Chalmers University of Technology\\
  Chalmersplatsen 1, 41296 Gothenburg, Sweden\\
  \texttt{burden@chalmers.se} \\
  \And
  Susanne Stenberg \\
  RISE Research Institutes of Sweden\\
  Lindholmspiren 7, 41756 Gothenburg, Sweden\\
  \texttt{susanne.stenberg@ri.se} \\
  \And
  Jan-Philipp Steghöfer \\
  XITASO GmbH\\
  Austraße 35, 86153 Augsburg, Germany\\
  \texttt{jan-philipp.steghoefer@xitaso.com} \\
  \And
  Niklas Kühl \\
  University of Bayreuth\\
  Wittelsbacherring 10, 95444 Bayreuth, Germany\\
  \texttt{niklas.kuehl@uni-bayreuth.de} \\
}

\begin{document}

\title{The Bidirectional Relationship Between XAI and Regulation: Operationalizing XAI for the AI Act}

\maketitle

\begin{abstract}
    The EU AI Act makes explainability urgent for high-risk AI systems, yet most XAI research focuses on technical metrics rather than regulatory compliance. Understanding how legal requirements reshape XAI method design is challenging: the AI Act regulates organizational relationships (providers, deployers) using legal terminology, specifies obligations without concrete technical requirements, and underrepresents end-users\,---\,the very stakeholders whose needs human-centered XAI addresses. As regulations emerge globally, human-centered XAI practitioners face both a challenge and an opportunity: regulations pull XAI research toward real-world deployment, while practitioners can actively shape how explainability enables compliance. This establishes a bidirectional relationship.
    
    Our contribution is threefold. First, we provide the first interdisciplinary analysis of XAI's role in the AI Act—conducted by a team comprising AI Act legal experts, ML engineers, and requirements engineers\,---\,on a real-world clinical decision support system. Second, we systematically align XAI stakeholder roles with AI Act legal responsibilities, revealing where explainability methods address regulatory requirements versus where additional measures are necessary. Third, we identify three key opportunities for human-centered XAI practitioners:
    actively defining their roles in regulatory implementation; making the user-to-affected-party relationship explicit where regulations address only provider-deployer obligations; and enabling compliance while building multi-level trust\,---\,from regulators to affected parties.
\end{abstract}
\vfill
Paper accepted at the Fourth World Conference on Explainable Artificial Intelligence, xAI 2026, Fortaleza, Brazil, July~1-3, 2026.

\pagebreak

%
\begin{acronym}
    \acro{ai}[AI]{artificial intelligence}
    \acro{ml}[ML]{machine learning}
    \acro{cnn}[CNN]{convolutional neural network}
    \acro{dnn}[DNN]{deep neural network}
    \acroplural{dnn}[DNNs]{deep neural networks}
    \acro{icu}[ICU]{intensive care unit}
    \acro{lstm}[LSTM]{long short-term memory}
    \acro{los}[LOS]{length of stay}
    \acro{nn}[NN]{neural network}
    \acroplural{nn}[NNs]{neural networks}
    \acro{mts}[MTS]{multivariate time series}
    \acro{rnn}[RNN]{recurrent neural network}
    \acro{xai}[XAI]{explainable artificial intelligence}
    \acro{cdss}[CDSS]{clinical decision support system}
    \acro{fria}[FRIA]{fundamental rights impact assessment}
\end{acronym}

\newpage
\section{Introduction}


After long and complicated negotiations between the European Parliament, the EU Commission, and the EU Member States, the AI Act entered into force in August 2024 \citep{aiact}. AI-based systems are now regulated under a comprehensive set of product safety rules for the first time. 
In the coming years, different parts of the EU AI Act will become applicable. Providers and deployers of AI systems need to learn how to apply the different articles of the act, while notified bodies and certification authorities will need to establish a process to assess these systems according to the regulation.
Many standardization efforts are already underway, but there will be significant uncertainty about how the AI Act has to be applied and how it relates to different aspects of AI currently under research.

Regulations such as the AI Act are especially important in high-risk \ac{ai} systems\,---\,such as those in healthcare\,---\,to ensure that they are safe and respect fundamental rights and values~\citep{van_kolfschooten_eu_2024}. The AI Act follows a risk-based approach: the higher the risk, the stricter the rule. For high-risk systems, it describes many obligations for the provider and the deployer of \ac{ai} systems.

\Acp{cdss}, which may fall under the definition of a high-risk \ac{ai} system in the EU AI Act, place the responsibility for the high-risk decision on the \emph{end-user} (typically, clinicians). 
However, the AI Act does not directly address end-user needs and instead primarily focuses on regulating the providers and deployers of \ac{ai} systems. One of the few mentions of end-users is the statement that high-risk AI systems must be designed so that humans can effectively oversee these systems. There are no technical requirements that prioritize the user's needs at the center of its goals~\citep{van_kolfschooten_eu_2024}.

In contrast, nowadays\footnote{This has not always been the case. A human-centered perspective on \ac{xai}\,---\,one that incorporates stakeholder needs, integrates insights from social sciences, and critically reflects on \ac{xai}'s capabilities and limitations\,---\,has only emerged in recent years. We focus exclusively on this approach (see \Cref{sec:xai_background})} more \ac{xai} researchers suggest putting end-users and their needs at the center of attention~\citep{langer_what_2021}. \ac{xai} methods provide descriptions of the outputs of opaque AI models, thereby satisfying the needs of the end-users. This is very helpful in high-risk applications such as healthcare. Previous work has started exploring the role that \ac{xai} plays w.r.t.\ ensuring the trustworthiness of high-risk applications under the EU AI Act~\citep{panigutti_role_2023,kiseleva_transparency_2022,longo_explainable_2024,esmaeilzadeh_challenges_2024,bibalLegalRequirementsExplainability2021}. \ac{xai} can contribute to implementing transparency and human oversight, which are key principles of the AI Act~\citep{panigutti_role_2023}. 

This creates both demands and opportunities for the \ac{xai} community, establishing a \emph{bidirectional relationship}: providers and deployers need transparency mechanisms to meet compliance obligations, while \ac{xai} practitioners have the tools to actively represent end-user perspectives and shape how legal requirements translate into stakeholder-centered implementations. Taking this line of thinking further, we believe that \ac{xai}\,---\,if designed correctly\,---\,could play a crucial role in fulfilling the obligations of deployers and providers of AI systems. However, we acknowledge that \ac{xai} alone cannot fulfill all obligations under the \ac{ai} Act. Full compliance requires additional measures such as the development of harmonized standards, risk management frameworks, and legal assessments, which are beyond the scope of XAI methods. 

Therefore, we ask:

\textbf{Research Question:} How can Explainable AI bridge the gap between end-user needs and requirements of the AI Act?

To answer the research question, we analyze the relationship between the end-user's \ac{xai} needs and the AI Act implications from a legal and technical perspective. We do this by using a concrete, \ac{ai}-based \ac{cdss}. This real-world application is currently under development in collaboration with clinicians and academic partners. The \ac{cdss} helps \ac{icu} coordinators who make choices about which patients can be moved out of the \ac{icu} with predicted key indicators such as the remaining \ac{los}. It is important to note that clinicians can use the predictions to make a decision in our \ac{cdss}\,---\,the system never decides on its own. To overcome acceptance and trust issues, we use \ac{xai} methods to provide human-understandable explanations of the system.
Research on the AI Act and research on \ac{xai} are often conducted independently, lacking interdisciplinary collaboration~\citep{bexAILawTransdisciplinary2025,gal_bridging_2023}. By providing a joint understanding of the role of \ac{xai} in the \ac{ai} Act from an interdisciplinary perspective representing various stakeholder needs, we\ldots
\begin{enumerate}
    \item provide an analysis of the implications of the AI Act in relation to adopting AI in healthcare (see \Cref{sec:act_understanding});
    \item describe how the needs of the \ac{cdss} stakeholders can be captured through XAI (see \Cref{sec:applicability-xai-exemplar}); and
    \item assess the similarities and contrasts between the legal requirements of the AI Act and the stakeholder desiderata of XAI (see \Cref{sec:reflections-xai-aia}).
\end{enumerate}
Our work is the first that analyzes the role of \ac{xai} in the \ac{ai} Act in an interdisciplinary team, on a real-world \ac{cdss} that takes user-centered needs into account and contributes to the state-of-the-art by mapping obligations of the EU AI Act with the obligations of stakeholders and how to fulfill them using \ac{xai}.
Through this analysis, we demonstrate how the \ac{xai} community can move from passive observers to active participants in regulatory implementation\,---\,positioning human-centered explainability as essential to compliance while ensuring that regulations pull \ac{xai} research toward real-world stakeholder needs.

\section{Related Work and Background}\label{sec:background-and-related-work}
In this section, we give an overview of the AI Act and \ac{xai}. We also discuss relevant related work.

\subsection{Background: The EU AI Act}\label{sec:the_ai_act}
In 2024, the European Union ratified the regulation of AI as technology \citep{aiact}. The new regulation was motivated by a need to create trust for AI with respect to product safety, health, and fundamental rights. The AI Act incorporates AI systems into the EU's existing framework for product safety. Certain AI systems require conformity assessment and subsequently carry the CE-mark as a sign of being a certified product, just like elevators, machinery, and medical devices.

The regulation also gives authorities an important role in guaranteeing the responsible development and usage of AI, as public sector bodies are to conduct market surveillance but also foster AI innovation (see e.g., \cite{regulatingtrust}). There are two central actors from the perspective of the regulation: the provider and the deployer of an AI system. These actors are obliged to take on responsibility for product safety.

\paragraph{CE-marking AI Systems.}
The conformity assessment for CE-marking AI systems can be done in two ways. The first is through self-assessment, where the provider of the system certifies that the requirements are met. This is also referred to as \emph{internal conformity assessment}. The second conformity assessment procedure involves a notified body, an organization that has been accredited to be competent in conformity assessment of certain products. The role of the notified body is to give a neutral assessment of the extent to which the requirements are met. The nature of the AI system will determine which process is to be applied \citep{conformityassessment}.

\paragraph{Obligations of the Provider.}
The provider is the organization creating the system, or as defined in the AI Act, \art{3}, a provider is ``a natural or legal person, public authority, agency or other body that develops an AI system or a general-purpose AI model or that has an AI system or a general-purpose AI model developed and places it on the market or puts the AI system into service under its own name or trademark, whether for payment or free of charge''.

Providers of high-risk AI systems must adhere to several obligations as outlined in \art{16-21}. They must ensure compliance with requirements in Section 2 of the AI Act (\art{9-15}) and have a quality management system per \art{17}. They need to maintain documentation and keep logs generated by their high-risk AI systems. The systems must undergo a conformity assessment, have a CE-mark, and be registered in the Commission's database. Providers also have to cooperate with authorities and take corrective measures if the system fails to comply with the AI Act.

The provider needs to perform risk management as outlined in \art{9}, which involves analyzing risks and developing mitigation strategies for the AI system. Transparency requirements in \art{13} ensure that operations are clear enough for deployers to interpret the system's output and use it appropriately. Human oversight, as specified in \art{14}, requires that the person operating the AI system understands the relevance of both input and output in the context of specific activities. Other technical requirements include data governance (\art{10}), technical documentation (\art{11}), record keeping (\art{12}), accuracy, robustness, cyber-security (\art{15}), and quality management~(\art{17}).
There will be harmonized standards issued by the European standardization organizations CEN and CENELEC that further detail the requirements in \art{9 to 15 and 17}. An overview of the standardization effort can be found on the two organizations' joint technical committee's website~\citep{europeancommitteeforstandardizationCENCENCLC2025}.

\paragraph{Obligations of the Deployer.}
According to the definition in \art{3}, a deployer is ``a natural or legal person, public authority, agency or other body using an AI system under its authority except where the AI system is used in the course of a personal non-professional activity''.

The obligations for deployers of high-risk AI systems, as outlined in \art{26}, include taking technical and organizational measures to ensure systems are used according to instructions. If a risk is identified, deployers must promptly inform the provider and relevant authorities and suspend system use. Furthermore, deployers must assign human oversight to competent individuals with necessary training and authority. Deployers controlling input data must ensure it is relevant and representative in view of the intended purpose of the system. They are responsible for maintaining logs appropriately and informing individuals when decisions affecting them involve high-risk AI systems. Additionally, deployers must cooperate with competent authorities in actions related to the AI system to implement the regulation.

\subsection{Background: Explainable AI}\label{sec:xai_background}
As indicated, current research on \ac{xai} provides a new perspective on the relationship between an \ac{ai} system and its affected parties. Therefore, we provide a brief nomenclature of relevant key concepts and terms. We define the key terms related to \ac{xai}, examine the various stakeholder terminologies, and clarify a common terminology of \ac{xai}\footnote{We acknowledge that terms like \textit{interpretability}, \textit{explainability}, \textit{transparency}, or \textit{trust} are used differently across disciplines. In this paper, we adopt definitions from the technical \ac{xai} literature to guide system design and evaluation. Our focus is on their operational meaning in the context of \ac{ai} Act compliance, rather than engaging in broader philosophical debates.} methods.

\paragraph{Scope of \ac{xai} in This Work.} Broad \ac{xai} goals such as trustworthiness, informativeness, and fairness have been widely studied~\citep{amparoreTrustNotTrust2021,deckMappingPotentialExplainable2024,deckCriticalSurveyFairness2024,hoffman_measures_2023,barredo_arrieta_explainable_2020}, but the area of study have recently shifted to focus more on individual stakeholders~\citep{uthPersonalizingExplanationsAIbased2026,longo_explainable_2024,biecek2024position}, their desiderata~\citep{langer_what_2021}, and expertise levels~\citep{mohseni_multidisciplinary_2021}. We explicitly focus on this \textbf{human-centered XAI}\,---\,designed with end-user comprehension and stakeholder desiderata in mind. When we refer to ``XAI'' or ``the XAI community'' throughout this work, we address researchers and practitioners working within this paradigm. This scoping is essential because only human-centered XAI can meaningfully bridge stakeholder needs and AI Act compliance requirements.

\paragraph{Explainability and Transparency.}
The terms \emph{explainability}, \emph{interpretability}, and \emph{transparency} are often used interchangeably in \ac{xai} literature~\citep{al-ansari_user-centered_2024,renftle_what_2024,hoffman_measures_2023,barredo_arrieta_explainable_2020}. The AI Act uses \emph{transparency} broadly to encompass technical documentation, system capabilities, and information enabling deployers to interpret the system's output and use it appropriately (\art{13}). We distinguish between \emph{explainability} as a desired property\,---\,the degree to which a human can understand the reasons behind a system's decisions~\citep{barredo_arrieta_explainable_2020}\,---\,and \textit{\ac{xai}} as methods that provide human-understandable explanations for predictions made by complex AI systems~\citep{wang_designing_2019,barredo_arrieta_explainable_2020}.

\paragraph{Stakeholder Terminology.}
\ac{xai} success relies on satisfying diverse stakeholder desiderata~\citep{langer_what_2021,fairhurstLegalRequirementsTrust2026}. \cite{langer_what_2021} categorize stakeholders into five classes: \textit{users}, \textit{developers}, \textit{affected parties}, \textit{deployers}, and \textit{regulators}. \cite{mohseni_multidisciplinary_2021} further divide end-users by expertise: \textit{Novice Users} (daily AI users with little ML knowledge), \textit{Data Experts} (data scientists and domain experts using ML for analysis), and \textit{AI Experts} (ML scientists and engineers implementing ML algorithms and XAI methods).


\paragraph{XAI Method Taxonomy.}
We adopt \cite{speith_review_2022} taxonomy to guide \ac{xai} selection for our CDSS (see \Cref{sec:design-choice}). \emph{Stage}: ante-hoc methods (inherently interpretable models) versus post-hoc methods (explain opaque models with independent algorithms). \emph{Applicability}: model-specific (using internal structure) versus model-agnostic (applicable to any model). \emph{Scope}: local explanations (specific input-output relations) versus global explanations (overall model behavior). Other dimensions include \emph{input data type}, \emph{output format}, and \emph{problem type}.

\subsection{Related Work}
Several studies have already addressed the question of how \ac{ai} can be applied in healthcare, incorporating ethical and legal considerations \citep{aboy_navigating_2024,porsdam_mann_eu_2024,karimian_ethical_2022,kiseleva_transparency_2022}. The function of \ac{xai} in fostering transparency and human oversight has also been the subject of considerable debate~\citep{panigutti_role_2023}.
Other works have dealt with the role of XAI in the AI Act (e.g., \citep{gal_bridging_2023,juliussenRightExplanationGDPR2025,metikosRightExplanationPractice2025}) or the GDPR (e.g., \citep{juliussenRightExplanationGDPR2025,metikosRightExplanationPractice2025}).
Nevertheless, \textit{how} \ac{xai} can be effectively utilized to fulfill the obligations of stakeholders as defined in the AI Act remains unexplored. \Cref{tab:related_work_comparison} provides an overview of the related work and indicates how we position our own contribution.

\begin{table*}[t]
    \caption{Related work can be divided into work that explores the role of \ac{xai} in the AI Act and work that analyzes the AI Act in healthcare from \ac{xai} or user perspectives. Our work extends these studies by combining the user-centric view with the \ac{xai} view on the AI Act in a real-world healthcare scenario.}
    \label{tab:related_work_comparison}
    \centering
    \begin{tabular}{lccccc}
        \toprule
        Paper                            & AI Act    & XAI       & Healthcare & Real World & User-Centric \\
        \midrule
        \cite{juliussenRightExplanationGDPR2025}          & \ding{51} & \ding{51} &            &            &              \\
        \cite{buttaboniRegulatoryTaxonomyAI2026}      & \ding{51} & \ding{51} &            &   &              \\
        \cite{gal_bridging_2023}        & \ding{51} & \ding{51} &            &            &              \\
        \cite{pavlidis_unlocking_2024}  & \ding{51} & \ding{51} &            &            &              \\
        \cite{fresz_how_2024}          & \ding{51} & \ding{51} &            &            &              \\
        \cite{onitiu_limits_2023}       & \ding{51} & \ding{51} & \ding{51}  &            &              \\
        \cite{solaiman_regulating_2024} & \ding{51} &           & \ding{51}  &            & \ding{51}    \\
        \cite{van_kolfschooten_eu_2024} & \ding{51} &           & \ding{51}  &            & \ding{51}    \\
        \cite{metikosRightExplanationPractice2025}          & \ding{51} & \ding{51} &            &  \ding{51}   &  \\
        \cite{panigutti_role_2023}      & \ding{51} & \ding{51} &            & \ding{51}  &              \\
        \midrule
        \textbf{Our Work}                & \ding{51} & \ding{51} & \ding{51}  & \ding{51}  & \ding{51}    \\
        \bottomrule
    \end{tabular}
\end{table*}

Several studies explore the role of \ac{xai} in the AI Act from conceptual and legal perspectives. Juliussen~\citep{juliussenRightExplanationGDPR2025} provides a comprehensive analysis of the right to explanation under both GDPR and the AI Act, examining \ac{xai} methods (intrinsic and post-hoc) in the context of automated decision-making. Buttaboni and Floridi~\citep{buttaboniRegulatoryTaxonomyAI2026} propose a regulatory taxonomy that distinguishes transparency, traceability, interpretability, and explainability as layered and interdependent dimensions of AI opacity, providing a conceptual framework for legal interpretation and compliance. \cite{gal_bridging_2023} explore the transparency gap between the AI Act and \ac{xai} understanding, emphasizing that transparency requirements are broad and should not be seen as concrete engineering instructions; \ac{xai} contributes to transparency but should not be viewed as a standalone solution. \cite{pavlidis_unlocking_2024} examines \ac{xai} as a mechanism to align AI Act objectives with end-user preferences, focusing on standard setting, oversight, and enforcement. \cite{fresz_how_2024} review legal requirements for explainability and derive requirements that \ac{xai} methods need to fulfill. Our work extends these studies by elaborating the role of \ac{xai} in a real-world healthcare scenario and incorporating end-user needs.

Within healthcare, Onitiu~\citep{onitiu_limits_2023} translates technical safeguards in \art{13-14} for medical diagnostic systems, emphasizing that post-hoc \ac{xai} methods must harmonize human expertise with patient-oriented values. Solaiman and Malik~\citep{solaiman_regulating_2024} explore the AI Act in healthcare with a user-centric focus, examining how the doctor-patient relationship has evolved to one where patients are active participants in their healthcare decisions, and how \ac{ai} and \ac{xai} can enhance patient autonomy by providing more information to both clinicians and patients. Van Kolfschooten and Van Oirschot~\citep{van_kolfschooten_eu_2024} analyze AI Act implications for healthcare stakeholders, providing implementation guidelines and highlighting the importance of collaboration with \ac{ai} developers, healthcare professionals, patient communities, and regulators. In contrast to these studies, our approach combines user-centric and \ac{xai} perspectives through an analysis of healthcare stakeholders from both AI Act and \ac{xai} viewpoints in a real-world scenario.

Real-world applications are examined by Metikoš and Ausloos~\citep{metikosRightExplanationPractice2025}, who analyze GDPR case law on the right to explanation across European judicial bodies and data protection authorities, providing insights into how the AI Act's right to explanation should be understood in practice. \cite{panigutti_role_2023} conduct an interdisciplinary analysis focusing on \art{13 and 14}, presenting a use case involving an AI-based proctoring system without \ac{xai} tools. In contrast, our study extends beyond \art{13 and 14} to examine other articles that establish obligations for healthcare stakeholders in a real-world healthcare scenario with integrated user needs.

\section{Methodology}
\label{sec:methodology}

To answer our research question, we apply a two-pronged methodology within an interdisciplinary team. This section first describes the team and the distribution of responsibilities and incentives. We then describe the \ac{cdss}. That is followed by the two prongs of the approach: the legal analysis of the AI Act and the XAI analysis, both in relation to the \ac{cdss}. We then show how the two prongs merge.

\subsection{The Interdisciplinary Team}
The team behind this contribution consists of five individuals representing diverse expertise. Two individuals from the company developing the \ac{cdss} bring backgrounds in \textit{requirements engineering}, \textit{empirical software engineering}, and \textit{human-centered \ac{xai}} implementation. They organized the stakeholder workshops (see \Cref{sec:act_understanding}) and maintain ongoing interdisciplinary exchange with clinicians within a publicly funded research project.
Two individuals from a research institute contributed \textit{legal} and \textit{technical AI Act} expertise: one with experience as an appeals court judge and involvement in governmental legislative investigations, the other with a PhD in computer science. Together, they assisted the Government's Office during AI Act negotiations and implementation, and currently advise notified bodies on conformity assessment procedures and AI Act applicability.
The fifth member is a university professor for \textit{information systems and \ac{ai}}, serving as an external auditor to assess the validity and scientific soundness of the contribution from both technical and socio-technical perspectives.

\subsection{The Clinical Decision Support System}\label{sec:the_exemplar} 
The \ac{ai}-based \ac{cdss} is designed to help \ac{icu} coordinators manage bed occupancy. When the \ac{icu} reaches capacity and new patients require intensive care, coordinators must decide which patients can be moved to another ward and which must remain in the \ac{icu}. Typically, such decisions rely on extensive clinical experience and a clinician's intuitive sense of whether a patient is stable enough to be transferred. However, when these decisions fall to assistant doctors who lack the same depth of experience, the quality of the decision-making process may be compromised.

This is where the \ac{cdss} comes into play. It supports clinicians by predicting key indicators that guide the transfer decision\,---\,such as remaining \ac{los}, mortality probability, and readmission probability\,---\,using vital signs like heart rate and oxygen saturation as well as indicators such as the need for ventilation or the presence of an infection. When the AI's predictions conflict with a clinician's intuition, the interpretability of the system is crucial for reconsidering and validating the final decision. To generate these predictions, a \ac{dnn} is trained on time-series patient data spanning the years 2017 to 2024.
Notably, the system does not suggest which patients to move, but rather shows the predictions to help clinicians make their own decisions.


\subsection{Legal Analysis}
\label{sec:methodology:legal-analysis}
The legal analysis consists of mapping the different articles of the AI Act to the system under investigation. Which technical and organizational requirements does the system imply? Who will be responsible for the fulfillment of the requirements? The details of the technical requirements are not as important for the analysis as their implications for the different roles. For the legal analysis, e.g., it is more important to know who is responsible for which aspects of data management (input data, validation data, etc.), than how to assess the relevance or completeness of data. As analyses emerge, we visualize the concepts and their relationships since text has the drawback of enforcing a linear representation (see e.g. \Cref{fig:xai-aiact-roles-responsibilities-a}). The primary outcome of the legal analysis is given in \Cref{sec:act_understanding}.

At the same time, we make decisions on what is in and out of scope. This is necessary since the AI Act consists of 113 articles that define roles, responsibilities, technical requirements, and so on. That is complemented by 13 annexes that supply further detail to the articles. To motivate and help with the interpretation of the articles and the annexes, there are 180 recitals. Knowing what to include in an analysis is therefore an analysis in itself and something that is made easier as more is known about the system under consideration. In the case of this contribution, we settle on focusing primarily on the responsibilities of the roles of the provider and deployer, leaving out the content related to what happens if a system does not comply. We also keep the level of detail abstract in terms of technical requirements as going into detail would yield enough content for multiple academic contributions per article of the AI Act.


\subsection{Requirements Engineering Activities}

To understand the requirements of the clinicians who are going to use the \ac{cdss}, we conduct a workshop in which we collect the most important functional and quality requirements. The workshop is modeled after the Design Thinking method~\citep{ferreira2019design}, but much abbreviated. Participants were clinicians and ML researchers, with a requirements engineer moderating the workshop.

We engage in brainstorming sessions in which we map out the processes in the \ac{icu} and identify the most relevant user stories. These are then transformed into a user journey using user journey mapping~\citep{milicic2012towards} to capture the requirements on the system implied by these processes. A scoping exercise then allows us to narrow down which aspects of the process can be supported by the system and thus define the scope of the \ac{cdss}.

To refine the requirements and the desiderata for the \ac{xai} methods in particular, we conduct several interviews with clinicians. We also perform contextual inquiry~\citep{bednar2009contextual} where we follow a clinician on their rounds in the \ac{icu} to better understand how the process we elicit works in practice and which factors they consider before deciding which patient to move out of the \ac{icu}.

\subsection{Stakeholder Mapping and XAI Analysis}
The next step of analysis is to identify the relevant stakeholder groups. We follow the stakeholder mapping approach outlined by \cite{langer_what_2021} and involve the \ac{ai} Act and \ac{xai} experts. Stakeholders identified during the Design Thinking Workshop are mapped, we summarize their specific desiderata, and expand on them in regular project meetings and exchanges. For the choice of \ac{xai} methods, the categorization by \cite{speith_review_2022} is used. The \ac{xai} expert analyzes the problem and selects common methods that align with the problem types and relevant requirements of the \ac{cdss}, as identified in the requirements engineering workshops.

\subsection{The Merging of the two Approaches}\label{sec:merging_approaches}
The legal and the \ac{xai} analysis are conducted in parallel so that they benefit from insights from the other approach. This takes place over three different phases. Phase one is a one-day sprint (March 2024) where the AI Act is introduced to the company and a subsequent workshop starts to map out the high-level applicability towards the \ac{cdss}. Phase two occurs six months later (September 2024) to share new insights in a second workshop. The third phase runs from December 2024 to January 2025 and is conducted as six weekly meetings with the ambition to document the combined analysis and the reasoning. Between the meetings of the third phase, each individual contributes to the shared document, the meetings discuss the combined effort, and changes are made as new insights are gained. The outcome in this contribution represents the final analyses.

\subsection{Validity Strategies}\label{sec:valifdity}
Following Lincoln and Guba~\citep{LincolnGuba1985}, we structure the validity strategies along  \textit{Credibility} (establishing an analysis true to the data and context);
\textit{Transferability} (ensuring that the findings were applicable beyond the context);
\textit{Dependability} (enabling repetition of our contribution); and \textit{Confirmability} (limiting the impact of our own incentives and bias).

As seen in \Cref{sec:merging_approaches}, we apply a methodology that allows for triangulation of methodologies applied separately by the individuals of the team, over time (Credibility and Confirmability). We also show how our analysis goes beyond a specific AI system in Sections~\ref{sec:reflections-xai-aia} and \ref{sec:discussion_future_research} (Transferability). The ambition is that it is possible to replicate the study as we describe the team behind the contribution and its incentives, as well as the applied methodologies for collecting and analyzing data (Dependability and Confirmability).

\section{Applicability of the AI Act to the CDSS}\label{sec:act_understanding}
As the first step of our overall analysis and as described in \Cref{sec:methodology:legal-analysis}, we analyze the \ac{cdss} in the context of the EU AI Act and establish whether the act is applicable to the case of the \ac{cdss} system. To do this, we need to answer three questions:
\begin{itemize}
    \item Does the system fall within the legal scope? (\art{2})
    \item Does the system fit the legal definition of an AI system? (\art{3 (1)})
    \item Does the system need to be CE-marked? (\art{6})
\end{itemize}

\paragraph{Scope.} The AI Act has a number of clauses defining the scope of the regulation. For instance, it ``does not apply to any research, testing or development activity regarding AI systems or AI models prior to their being placed on the market or put into service'' or systems used in a purely personal, non-professional activity. After going through the scope as defined in \art{2}, our assessment is that the AI system under investigation is within scope of the AI Act as it is to be put into use in a clinical context.

\paragraph{AI system.} An AI system in the AI Act is defined as ``a machine-based system that is designed to operate with varying levels of autonomy and that may exhibit adaptiveness after deployment, and that, for explicit or implicit objectives, infers, from the input it receives, how to generate outputs such as predictions, content, recommendations, or decisions that can influence physical or virtual environments''. A key concept is \textit{infer}, which can be done by machine-learning approaches but through knowledge- or logic-based technologies. However, ``systems that are based on the rules defined solely by natural persons to automatically execute operations'' should not be seen as AI systems (recital 12). Our assessment is that the AI system investigated is within the scope of the definition of an AI system as it, from received input, generates a prediction that, if the clinician approves, will influence where and by whom the patient will be treated.

\paragraph{CE-mark.} There are two scenarios where the AI system needs to be CE-marked: the first covers clinical systems that fall under EU regulation 2017/745 on medical devices and subsequently need to undergo a conformity assessment procedure involving a notified body (\art{6.1} of the AI Act); the second is for AI systems that are listed in Annex III of the AI Act. In the latter case, point 5.a refers to ``AI systems intended [\ldots] to evaluate the eligibility of natural persons for [\ldots] healthcare services, as well as to grant, reduce, revoke, or reclaim such [\ldots] services''. 

Our assessment is that, while it is unclear if the \ac{cdss} is a medical device or not, it does not fulfill the criteria to fall into a risk category mandating the involvement of a notified body, according to regulation 2017/745 on medical devices. However, in preparing for the regulated case scenario, we do assess that the AI system under investigation is covered by the list in Annex III. Some systems under Annex III can be exempt from CE-marking if they are solely used for preparatory purposes (\art{6.3a.d}), but since our system conducts profiling as defined in General Data Protection Regulation (GDPR), we cannot apply this exempt (\art{6.3}).

To summarize, the AI system needs to be CE-marked through an internal conformity assessment procedure in accordance with the AI Act, i.e., without involving a notified body (Annex IV of the AI Act). This is the obligation of the provider. In practice, this means that the provider needs to fulfill several obligations, as described in \Cref{sec:the_ai_act}. This includes fulfilling technical criteria (\art{9-15}) and having a quality management system (\art{17}).

That is mirrored by the obligations of the deployer (\art{26}), described in \Cref{sec:the_ai_act}. Since the system will be deployed by a legal person governed by public law (or by private entities providing public services), there is another obligation: a \ac{fria} shall be conducted by the deployer (\art{27}). How to do this will be detailed in a harmonized standard.

We use our findings from this legal analysis in \Cref{sec:reflections-xai-aia} as the foundation for contrasting XAI and the AI Act.

\section{Choosing the right XAI Method}\label{sec:applicability-xai-exemplar}
Choosing the right \ac{xai} method is important to ensure that stakeholder needs are satisfied~\citep{schoonderwoerd_human-centered_2021}. Therefore, we analyze relevant stakeholders and elaborate their desiderata (see \Cref{sec:our_stakeholder}). This information helps us make well-founded choices about the \ac{xai} method in the implementation of the \ac{cdss} (see \Cref{sec:design-choice}).

\subsection{Our Stakeholders and their Desiderata}\label{sec:our_stakeholder}
\begin{table*}[!t]
    \centering
    \caption{Overview of the real-world stakeholders and desiderata mapping.}
    \label{tab:desiderata_mapping}
    \resizebox{\textwidth}{!}{%
        \begin{tabular}{@{}P{3 cm}P{3.5cm}P{4.3cm}P{4.4cm}@{}}
            \toprule
            \textbf{XAI Stakeholder Roles} & \textbf{AI Act Stakeholder Roles}                  & \textbf{Clinical Role}      & \textbf{Desiderata}          \\
            \midrule
            \textbf{(System) Developer}    & Provider                                           & ML Researcher and Engineers & Reliability, Performance     \\
            \textbf{Deployer}              & Deployer                                           & Hospital Management         & Acceptance, Legal Compliance \\
            \textbf{User}                  & Natural Person                                     & Clinicians                  & Interpretability, Usability, \\
                                           &                                                    &                             & Trust/ Appropriate Reliance  \\
            \textbf{Affected Parties}      & Natural Person(s)                                  & Patients                    & Correctness, Fairness        \\
            \textbf{Regulator}             & AI Office, AI Board, Competent Authorities, \ldots & Competent authorities       & Compliance                   \\
            \bottomrule
        \end{tabular}
    }
\end{table*}

The first group of stakeholders are researchers and engineers specializing in \ac{ml}, who are categorized as \emph{developers} and identified as providers under the provisions of the AI Act. Their primary requirement for the \ac{ai} system is the achievement of high performance and reliability. In order to fulfill this requirement, they concentrate on quality measures such as accuracy and robustness, which are of the highest importance for the practical implementation and success of the model in real-world applications.
This will increase end-users' acceptance of our \ac{ai} system, which is a key factor for hospital management.

As the \emph{deployer} of the \ac{ai} system, the hospital management is interested in the acceptance of the \ac{ai} system by the clinicians. It needs to be a reliable source of support in high-stakes decision-making processes. Furthermore, the \ac{ai} system must comply with a range of regulations, such as the AI Act, to ensure its appropriateness in the hospital.

Clinicians are the primary \emph{users} of the AI system in \ac{cdss}, as they are the only individuals who directly interact with the AI outputs and ultimately make the final decisions. Most of these clinicians are novice users; as defined in \Cref{sec:xai_background}, the \ac{ai} system must be accessible and user-friendly. This translates into the desideratum interpretability, which involves providing comprehensible, plausible, and correct explanations for the AI system. This interpretability allows clinicians to question AI predictions rather than simply accepting them, thereby fostering trust and appropriate reliance~\citep{schemmer_appropriate_2023}. Clinicians should trust the AI system when it is correct and appropriately rely on their judgment when the AI system is incorrect. This is most important when the \ac{ai} system makes a false prediction that negatively impacts the patient, e.g., the \ac{ai} system falsely predicts a short remaining \ac{los} for the patients, although the patient is very sick and will have to stay longer. In this situation, the clinicians must rely on their medical expertise.

Patients, although not directly interacting with the AI system, are significantly impacted by the decisions made by clinicians using the AI system and thus an ``\emph{affected party}'' according to the definition in \Cref{sec:xai_background}. Patients have a strong need for correct and fair decision-making. Fairness in AI-driven decisions is a critical requirement, as it directly affects patient outcomes and trust in the healthcare system. The \emph{regulator} corresponds to the competent authority, which can be mapped to different \ac{ai} Act stakeholder roles, e.g., \ac{ai} Office or \ac{ai} Board.

\subsection{Our XAI Method Choices}\label{sec:design-choice}
After we clarified the relevant stakeholders and their desiderata, we describe the concrete \ac{xai} method choices we made for our \ac{cdss}. Categories like input data and problem type are pre-determined by the problem the application addresses.

The \ac{ai} system is based on a complex \ac{dnn} that processes multivariate time series data, including patient vital signs, to predict the remaining \ac{los} of an individual patient. While \acp{dnn} provide great performance by learning complex relationships in data, they are not intrinsically interpretable and lack transparency~\citep{barnes_evaluating_2024}. Therefore, we employ post-hoc \ac{xai} methods to provide insights into this black-box model. To ensure a reliable and permanent \ac{ai} model, we dynamically retrain and update the \ac{ai} model to avoid data drifts. Some changes in the \ac{ai} model might affect the \ac{ai} architecture, making model-specific \ac{xai} methods potentially incompatible. For this reason, we selected model-agnostic \ac{xai} methods that can be applied to any underlying \ac{ai} model based on a neural network structure.

The primary goal of our \ac{xai} method is to satisfy our end users' desiderata and thus target interpretability, usability, and trust or appropriate reliance. Our \ac{xai} method focuses on a local \emph{scope}, enabling clinicians to understand specific predictions by interpreting relationships between input data, such as patient vital signs, and predicted outcomes, like the estimated remaining \ac{los}. This interpretability is crucial to ensure appropriate reliance in clinical decision-making. To ensure that the output of the \ac{xai} method is understood accordingly, we use a combination of different output formats, such as visual and numerical information, providing appropriate representations for users with different desiderata and levels of expertise.
\begin{figure}[t]
    \centering
    \includegraphics[width=0.7\linewidth]{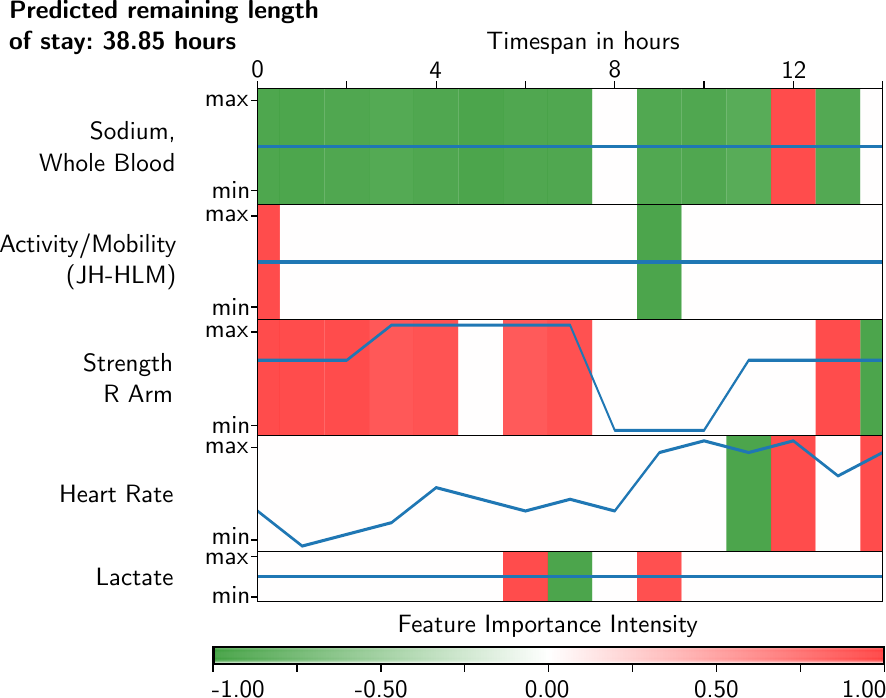}
    \caption{Example Heatmap Explanation with SHAP. Shows the five most important features in descending order of relevance over a period of 14 hours. The color gradient indicates whether the feature value at a given timestamp has a positive (red) or negative (green) impact on the predicted \ac{los}.}
    \label{fig:shap-explanation}
\end{figure}

Finally, the different choices in the categories allow us to choose between \ac{xai} methods such as SHAP~\citep{lundberg_unified_2017}, LIME~\citep{ribeiro_why_2016}, Integrated Gradient~\citep{sundararajan_axiomatic_2017}, or Saliency Maps~\citep{simonyan_deep_2014}. Functionally-grounded evaluation methods~\citep{doshi-velez_towards_2017,nauta_anecdotal_2023} that measure explanation properties such as complexity, faithfulness, or sensitivity help us to select the best \ac{xai} method~\citep{chenWhatMakesGood2024}.

For example, \Cref{fig:shap-explanation} shows a concrete heatmap explanation highlighting the importance of features for all time series steps. Here, the clinician gets an overview of the five most important features over the last 15 hours of the patient in the \ac{icu}. The color gradient indicates if the specific feature value at the specific time stamp has a negative (green) or positive (red) effect on the predicted remaining \ac{los}. A short \ac{los} indicates a more positive state of health then a long \ac{los}. This explanation of the \ac{los} prediction assists the clinician in appropriately deciding whether the patient can be moved to another ward.


\section{Contrasting XAI with the AI Act}\label{sec:reflections-xai-aia}
By analyzing the same \ac{cdss} from the perspectives of \ac{xai} and the AI Act, we have created opportunities to reflect on what one perspective reveals about the other. This allows us to answer our research question.

The AI Act refers to stakeholder obligations and responsibilities, while human-centered \ac{xai} research focuses on stakeholder desiderata. Mapping these two perspectives is not a straightforward one-to-one mapping. Rather than exhaustively detailing the inter-relationships, we explored a thin slice of their intersection to show how \ac{xai} can fulfill the AI Act's legal requirements. Our main finding is that the user-centered implementation of XAI enables conformant representation. A schematic view of the analysis is found in diagrams of \Cref{fig:xai-aiact-roles-responsibilities}. This means we are explicitly expecting further research to complement and refute our findings (see KI1-KI3), both in scope and in conclusions, if \ac{xai} is chosen as a relevant way to comply with the requirements of the AI Act.

\subsection{Roles, Responsibilities and Needs}
\begin{figure}[tbp]
    \centering
    \begin{subfigure}{\textwidth}
        \centering
        \includegraphics[width=0.8\linewidth]{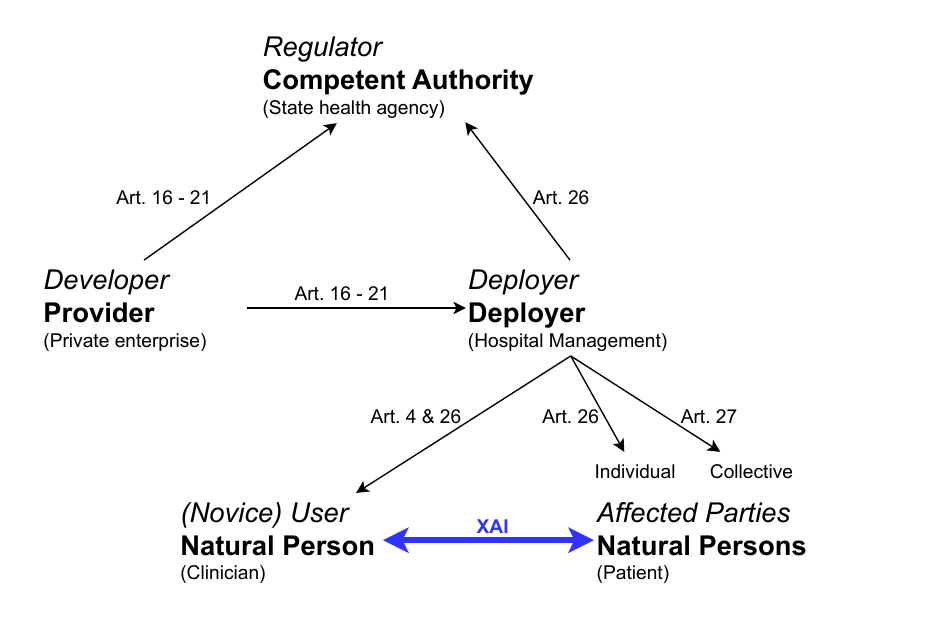} 
        \caption{The stakeholder roles and responsibilities when combining the AI Act with XAI.}
        \label{fig:xai-aiact-roles-responsibilities-a}
    \end{subfigure}
    \begin{subfigure}{\textwidth}
        \centering
        \includegraphics[width=0.8\linewidth]{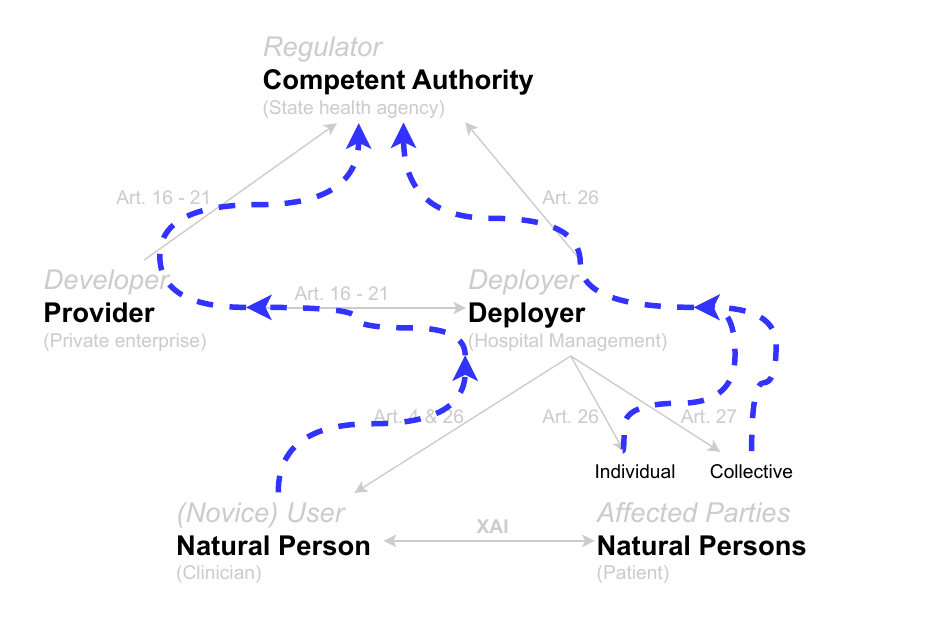} 
        \caption{Introducing XAI to comply with legal obligations will involve the entire stack of stakeholders.}
        \label{fig:xai-aiact-roles-responsibilities-b}
    \end{subfigure}
    \caption{Stakeholder roles and their responsibilities in XAI and the AI Act. \textit{Italics} denote XAI roles, \textbf{bold} denotes AI Act roles. Parenthesized roles are from the \ac{cdss}. An arrow from A to B denotes that A has a responsibility towards B.}
    \label{fig:xai-aiact-roles-responsibilities}
\end{figure}
While there is an overlap in the main stakeholder roles introduced by \cite{langer_what_2021} and the AI Act, there are also differences. Where \cite{langer_what_2021} taxonomy includes a single regulator role, the AI Act has multiple roles. There is the national authority assigned to conduct market surveillance of AI systems, the notified body that inspects to what extent an AI system, the developer's organization conforming to the requirements, and the EU Commission who have a say in how the AI Act is to be implemented and enforced. Even a court could serve as a regulator if the system developer and the national authority have different opinions. In the case of our \ac{cdss}, the system developer can even take the XAI role of regulator. This can seem counterintuitive, but it is one way of interpreting the conformity procedure based on self-assessment.

On the other hand, \cite{langer_what_2021} pay explicit attention to the end user and the affected parties, while the AI Act trusts the provider and the deployer to take care of those perspectives. As such, the stakeholder roles can be seen as lacking a common ground, or as complementing each other. That leaves ample room for the XAI community to define their stakeholder roles in relation to the responsibilities enforced by the AI Act.

Another observation is that while \cite{langer_what_2021} treat individuals ``as representative members of specified stakeholder classes'', our analysis mandates us to distinguish between affected parties as concrete individuals and as an abstract collective due to the responsibilities of the deployer. \art{26} in the AI Act defines requirements on the usage of the AI system on an individual level at run-time, while \art{27} poses requirements on performing risk assessments of the AI system in relation to its intended usage pre-launch in respect to the affected parties in general. This is visualized in \Cref{fig:xai-aiact-roles-responsibilities-a} as two different responsibilities (arrows) between the deployer and the affected parties.


\begin{tcolorbox}[colback=gray!20, colframe=gray!80, title=Key Insight (KI1)]
XAI practitioners must define their roles across AI Act responsibilities\,---\,moving from research observers to active participants in regulatory implementation.
\end{tcolorbox}


\subsection{Bridging the AI Act's Gap} 
The AI Act deliberately excludes the will of the affected parties (in our case, individual patients), focusing instead on roles relevant to product safety. Both the AI Act and XAI have roles that act on behalf of the affected party. In the AI act, the affected party lacks the ability, mandate, or obligation to change the functionality of the AI system. In contrast, XAI emphasizes the needs of affected parties in relation to the user, potentially bridging the gap between end users and affected parties, which the AI Act perspective lacks (highlighted as blue arrow in \Cref{fig:xai-aiact-roles-responsibilities-a}).

Notably, the AI Act does not explicitly define stakeholder roles, instead referring to ``relevant stakeholders'', ``external stakeholders'', ``multi-stakeholder'' etc. It is therefore our analysis that defines Deployer, Competent Authority, and Provider as stakeholder roles. Regarding the equivalence to the \ac{xai} roles of User and Affected parties, the AI Act refers to the former as ``staff'' (the collective that shall be informed, trained and have responsibilities regarding AI systems according to \art{17} on quality management systems), ``affected workers'' (those who will use the CE-marked system in relation to \art{26} and the impact on individuals' rights) and ``natural persons'' (those who have the adequate training for using a CE-marked AI system, \art{26}). The latter role is referred to as ``categories of natural persons or groups of persons'' (impacted by CE-marked systems under \art{27}). As the AI Act lacks systematic terminology for stakeholder roles, we use \emph{natural person(s)} to emphasize human involvement (natural person is legal terminology for human) rather than legal persons (legal terminology for organizations, companies, authorities, etc.).

In contrast, stakeholder-oriented \ac{xai} methods explicitly consider the affected parties, for instance by personalizing explanations and thus directly addressing the perspectives of individuals affected by AI-based systems~\citep{uthPersonalizingExplanationsAIbased2026,quahPERSONALIZEDEXPLAINABILITYREQUIREMENTS2025}. By customizing explanations based on their characteristics~\citep{schneiderPERSONALIZEDEXPLANATIONMACHINE}, they become understandable across different levels of domain and \ac{ai} expertise. This shared understanding of AI explanations facilitates dialogue among and between different user groups and affected parties (e.g., clinicians and patients).

\begin{tcolorbox}[colback=gray!20, colframe=gray!80, title=Key Insight (KI2)]
XAI fills a critical gap in the AI Act: while regulations address provider-deployer obligations, XAI makes the user-to-affected-party relationship explicit and actionable.
\end{tcolorbox}

\subsection{From Regulator to Affected Parties --- and back}
The AI Act mandates respecting the fundamental rights of affected parties.
In our case, a \ac{fria} must be conducted before system deployment, requiring the deployer to assess the impact and risks to patients' fundamental rights (\art{27}, which relates to Fairness). This links back to the provider's obligations. The \ac{fria} uses information from the AI-system provider (\art{13}) and the provider's risk assessment (\art{9}). Technical documentation and logs help the provider conduct the \ac{fria} and assist the user in evaluating output relevance, impacting XAI desiderata like trust and acceptance.
However, the AI Act does not specify technical methods or thresholds.

This opens a research trajectory for the XAI community on how to achieve appropriate methods and reasonable cut-off points. As an example, how certain does a recommendation system need to be about key-indicators, such as the \ac{los}, that support the decision which patient is moved from the ward to make room for another patient? How do we calculate that certainty, and how is it represented to physicians, hospital management, regulators, and affected parties? For this to be relevant in a legal context, the future contributions of the XAI community need to be understood and accepted by the diverse group of actors that can take the stakeholder roles of deployer, developer and regulator (see \Cref{fig:xai-aiact-roles-responsibilities-b}). This aligns with \cite{pavlidis_unlocking_2024}, who emphasizes that \ac{xai} must operate in compliance settings to meet regulatory requirements. In our case, it led us to target interpretability, usability and trust as key needs to address.

The AI Act introduces additional operational activities like a data lifecycle and risk management aspects (e.g., the delegation of responsibilities), without specifying methods. Here, the XAI community has an opportunity to delimit what is within its interests to contribute towards, but also new aspects of XAI development and usage for future contributions.

\begin{tcolorbox}[colback=gray!20, colframe=gray!80, title=Key Insight (KI3)]
XAI simultaneously facilitates compliance with the AI Act by offering essential documentation for \ac{fria} and builds multi-level trust\,---\,from regulators to affected parties.
\end{tcolorbox}

\section{Discussion and future research}\label{sec:discussion_future_research} 


From the \ac{xai} perspective, the starting point of our analysis is the obligations of the deployer. The persons assigned the oversight or usage of the system need to have the right competence, authority, and support so that they can assess if the output is reasonable given the input and current state of the \ac{icu} (\art{26} of the AI Act, which relates to the XAI term of Interpretability). As described in \Cref{sec:the_exemplar}, the \ac{cdss} provides key indicator recommendations to support the clinical decision which patients can be moved to another ward. While XAI can provide measures for explaining how the recommendation came to be, it does not assess whether it is right or wrong. Operating the ward still requires human judgment to assess which recommendations to follow and when alternative routes are more appropriate (see, e.g., \citep{schoeffer_explanations_2024,gomez_human-ai_2025} on Human-AI Collaboration).



The design choices we propose for the AI system under consideration are influenced by the stakeholders' needs but also help us to fulfill the obligations of the AI Act, e.g.:
\begin{itemize}
    \item Risk Management: We employ local XAI methods such as SHAP that have the potential to enable clinicians to critically evaluate the predictions by the AI system. This is one way to mitigate the potential risk of the end-user relying uncritically on the AI. Therefore, XAI provides a risk mitigation measure as suggested by the AI Act in \art{9} (2d).
    \item Transparency: Our \ac{xai} method provides information that can help clinicians ensure an appropriate reliance on \ac{ai} systems, thereby ``enab[ling] deployers to interpret the system's output and use it appropriately'' (see \art{13}). Especially, finding the ``appropriate type and degree of transparency'' needs to involve stakeholders and take their individual desiderata into account. Transparency can thereby be seen as a precondition to \ac{xai}.
    \item Human Oversight: An appropriate and individualized \ac{xai} method, in our case including complex outputs like heatmaps, is crucial for ensuring that high-risk AI systems can be effectively monitored by humans during their operational period, as stated in the AI Act (see \art{14}).
\end{itemize}

The overall ambition of the AI Act is to create trust in AI as a technology. It does so by stating general requirements that need to be addressed concretely, both in relation to the intended purpose of the system but also to the involved actors. Here \ac{xai} can be of use. As detailed in \Cref{sec:our_stakeholder}, the selection of \ac{xai} method must be tailored to the stakeholders and their desiderata. The success of this approach depends on meeting the specific desiderata and the satisfaction of each end-user. Engaging relevant stakeholders in the selection of \ac{xai} method allows addressing their individual desiderata and an effective implementation of different elements of the AI Act.

Yet, there remains much to do.
Since the \ac{xai} method choices reflect what happens between the physician and the patient (novice user and affected party), they need to be percolated upwards to the hospital management (deployer) of the AI system and captured as requirements, consequently fulfilled by the system developer (developer) of the AI system. Finally, the design choices and their implementation need to be understood and accepted by the regulator (see \Cref{fig:xai-aiact-roles-responsibilities-b}). \ac{xai} can serve as an instrument to build trust at all levels in relation not only to AI as technology but to a specific AI system\,---\,the regulator trusts the AI system to be fit for purpose, the deployer and provider trust the AI system to be fit for purpose and the user and the affected parties trust the AI system to be fit for purpose. A similar reasoning is applicable for the usability of the AI system\,---\,it needs to be usable for both the end user and the deployer. We foresee a number of studies to further explore the relationship between the levels but also between \ac{xai} as a tool and the legal requirements of the AI Act and how they are further detailed in the upcoming harmonized standards.

Here, it is worth pointing to an emerging discussion within the academic community on the risks of introducing AI in the healthcare system, specifically in the relationship between clinician and patient and the liability concerns in relation to perceived mistreatment (see, e.g., \citep{solaiman_regulating_2024,Duffourc2023,Cartolovni2023}). XAI could both serve as a means to lessen the risk of litigation, but also, in itself, poses a risk if not implemented properly. This is exemplified by \cite{weller_transparency_2022}, who states that explanations can lead to unsubstantiated trust to the lay user, and hence have a negative impact on affected parties. If \ac{xai} is to realize its potential in relation to the AI Act it is important that it is not applied to ``XAI-wash'' unsuitable systems (see e.g., \citep{rudin_stop_2019} who criticizes the use of XAI methods that may not be suitable in high-stakes decisions which are nonetheless justified using XAI), but for a genuine wish to make AI a suitable technology for human needs.

\section{Conclusion}\label{sec:conclusion}
As the AI Act enters into force, AI-based systems are now regulated under comprehensive product safety rules. The \ac{ai} Act mandates organizational obligations while underrepresenting end-users\,---\,creating a critical opportunity for human-centered \ac{xai}. By analyzing a \ac{cdss} under development in relation to the \ac{ai} Act and by analyzing the stakeholders from different perspectives, we map stakeholders' desiderata to regulatory requirements, making the role of the (novice) user and the relation to affected parties explicit. We demonstrate how human-centered \ac{xai} can bridge the gap between legal compliance and stakeholder needs.

Our analysis reveals three key opportunities for the \ac{xai} community. First, human-centered \ac{xai} practitioners must actively define their roles across AI Act responsibilities\,---\,moving from research observers to active participants in regulatory implementation. Second, human-centered \ac{xai} fills a critical gap: while regulations address provider-deployer obligations, \ac{xai} makes the user-to-affected-party relationship explicit and actionable. Third, stakeholder-tailored \ac{xai} simultaneously enables regulatory compliance (\ac{fria}, documentation, risk management) and builds multi-level trust\,---\,from regulators to affected parties.

We call on the \ac{xai} community to seize this moment: while regulatory compliance requires multiple components\,---\,from harmonized standards to risk management frameworks\,---\,human-centered \ac{xai} plays a unique role in ensuring transparency mechanisms address genuine stakeholder needs. This \emph{bidirectional relationship} where regulations reshape \ac{xai} research priorities while \ac{xai} practitioners shape transparency implementation offers an opportunity to ensure that those affected by \ac{ai} systems and their regulation have their interests represented in regulatory practice.




\bibliographystyle{unsrtnat}
\bibliography{bib/references_act, bib/cas-refs}




\end{document}